# Partially-coherent waves with binominal coherence

MIKHAIL CHARNOTSKII

*mikhail.charnotski@gmail.com*

**Abstract:** Comprehensive analysis of non-diffracting optical waves with binominal two-point coherence function (BCF) is presented. This coherence function consist of two terms, each depending on either separation of points or central point. We established the necessary and sufficient conditions for non-negative definiteness of the binominal coherence functions. Spectral analysis, including calculations of the eigenvalues and eigenfunctions for the general BCF case is presented. We considered two simple BCF examples of and analyzed the conditions leading to the maximum irradiance contrast.

## 1. Introduction

We discuss a specific case of the two-point Coherence Function (CF), also known as cross-spectral density, [1], of optical waves that has the form of a sum of two terms depending on the separation vector and the center position of the of two points:

$$\langle u(\mathbf{R}+\mathbf{r}/2)u^*(\mathbf{R}-\mathbf{r}/2)\rangle = W(\mathbf{R},\mathbf{r}) = \Phi(\mathbf{r}) + \Psi(\mathbf{R}). \tag{1}$$

Here $u(\mathbf{r})$ is the complex amplitude of a random quasi-monochromatic optical wave at the point $\mathbf{r}$ in some plane orthogonal to the wave propagation direction, angular brackets indicate the averaging over the wave fluctuations. We call this type of coherence functions Binominal Coherence Function (BCF) as implied by the analytical form of Eq. (1). The first component $\Phi(\mathbf{r})$ has uniform irradiance, and the spatially uniform degree of coherence. The second component, designated as "bump" in [2], is spatially non-uniform, but does not represent an irradiance distribution of its own accord.

Any CF, including the BCF is Hermitian,

$$W(\mathbf{R},-\mathbf{r}) = W^*(\mathbf{R},\mathbf{r}), \tag{2}$$

and satisfies Non-Negative Definite (NND) condition:

$$S[f(\bullet)] = \iint d^2r_1 \iint d^2r_2 W\left(\frac{\mathbf{r}_1+\mathbf{r}_2}{2}, \mathbf{r}_1-\mathbf{r}_2\right) f(\mathbf{r}_1)f^*(\mathbf{r}_2) \geq 0, \tag{3}$$

which has to be valid for any square integrable probe function $f(\mathbf{r})$, [1].

BCF waves were introduced in [2], where it was stated that they represent a general non-diffracting solution of paraxial equation for propagation of the two-point CF.

$$\left(ik\frac{\partial}{\partial z} + \nabla_{\mathbf{R}}\nabla_{\mathbf{r}}\right)W(\mathbf{R},\mathbf{r},z) = 0. \tag{4}$$

This statement is not fully accurate, since, for example, CF in the form

$$W(\mathbf{R},\mathbf{r}) = W(X,Y,x,y) = F(X,y) + G(Y,x) \tag{5}$$

is also $z$ – independent solution of Eq. (4). However, it is possible to show that Eq. (5) is not NND. Author is unaware of any examples of non-diffracting CFs that satisfy NND condition other than one presented by Eq. (1).

Paper [2] presented an example of BCF where both components are zero-order Bessel functions,

$$W(\mathbf{R},\mathbf{r}) = J_0(\beta r) + \alpha J_0(\beta R), \tag{6}$$

and it was noted that this CF is NND if and only if $|\alpha| \leq 1$.

Propagation of BCF was considered in [3], where it was stated that the NND of the BCF is "guaranteed" when Fourier transforms of each component is nonnegative. No restriction on the relative amplitudes of components have been imposed. In the following comment, [4] it was shown on a simple example that this NND condition is not sufficient. The author's reply [5], consequently acknowledged that the uniform component needs to be "high enough" to guarantee the NND, without specifying any quantitative restrictions. It should be noted that condition $|\alpha| \leq 1$ in [2] implies that Fourier transform of the inhomogeneous component can be negative.

Here we present comprehensive analysis of non-diffracting BCF optical waves, and establish rigorous necessary and sufficient NND conditions for the general case of the homogeneous and inhomogeneous components. In Section 2 we derive these conditions directly from the NND inequality, Eq. (3). In Section 3 we calculate the eigenvalues and eigenfunctions of the BCF and verify the results of Section 2 based on the non-negative feature of the eigenvalues. Section 4 presents simple examples of the bona fide BCFs and samples of random fields with binominal coherence. Technical calculation details are presented in three Appendices.

## 2. Non-negative definiteness of BCF

It proved to be convenient to represent all functions used in the following derivations as sums of real, imaginary even and odd components. For example, the probe function in Eq. (3) is presented as

$$f(\mathbf{r}) = f_{R.E}(\mathbf{r}) + f_{R.O}(\mathbf{r}) + i f_{I.E}(\mathbf{r}) + i f_{I.O}(\mathbf{r}). \tag{7}$$

Here subscripts $R$, $I$, $E$, and $O$ indicate real, imaginary, even and odd parts. In particular, Eq. (2) suggests that

$$\Phi(\mathbf{r}) = \Phi_{R.E}(\mathbf{r}) + i\Phi_{I.O}(\mathbf{r}), \quad \Psi(\mathbf{R}) = \Psi_{R.E}(\mathbf{R}) + \Psi_{R.O}(\mathbf{R}). \tag{8}$$

We introduce Fourier transforms of $\Phi(\mathbf{r})$, $\Psi(\mathbf{R})$ and probe functions $f(\mathbf{r})$

$$\hat{\Phi}(\mathbf{P}) = \frac{1}{4\pi^2} \iint d^2 r \Phi(\mathbf{r}) \exp(-i\mathbf{P} \cdot \mathbf{r}) = \hat{\Phi}_{R.E}(\mathbf{P}) + \hat{\Phi}_{R.O}(\mathbf{P}),$$
$$\hat{\Psi}(\mathbf{p}) = \frac{1}{4\pi^2} \iint d^2 R \Psi(\mathbf{R}) \exp(-i\mathbf{p} \cdot \mathbf{R}) = \hat{\Psi}_{R.E}(\mathbf{p}) + i\hat{\Psi}_{I.O}(\mathbf{p}), \tag{9}$$
$$\hat{f}(\mathbf{K}) = \frac{1}{4\pi^2} \iint d^2 r f(\mathbf{r}) \exp(-i\mathbf{K} \cdot \mathbf{r}) = \hat{f}_{R.E}(\mathbf{K}) + \hat{f}_{R.O}(\mathbf{K}) + i\hat{f}_{I.E}(\mathbf{K}) + i\hat{f}_{I.O}(\mathbf{K}),$$

and use them in the NND inequality, Eq. (3) to present it as

$$S[f(\bullet)] = (2\pi)^4 \left[ \iint d^2 P \hat{\Phi}(-\mathbf{P}) |\hat{f}(\mathbf{P})|^2 + \iint d^2 p \hat{\Psi}(-\mathbf{p}) \hat{f}\left(\frac{\mathbf{p}}{2}\right) \hat{f}^*\left(-\frac{\mathbf{p}}{2}\right) \right] \geq 0 \tag{10}$$

for all probe functions.

For pure homogeneous CF, when $\hat{\Psi}(\mathbf{p}) \equiv 0$, the NND condition is fulfilled if and only if $\hat{\Phi}(\mathbf{P}) \geq 0$. This is classic Bochner's theorem [6]. However, contrary to the statement in [3], similar condition, $\hat{\Psi}(\mathbf{p}) \geq 0$, does not warrant the non-negative values for the second term in Eq. (10) or for the $S[f(\bullet)]$.

After accounting for the symmetries detailed in the right-hand parts of Eq. (9), and simple change of variables in the second term in Eq. (10), the NND functional simplifies to

$$S[f(\bullet)] = (2\pi)^4 \iint d^2 P \left[ \hat{\Phi}_{R.E}(\mathbf{P}) A_f(\mathbf{P}) - 2\hat{\Phi}_{R.O}(\mathbf{P}) B_f(\mathbf{P}) \right.$$
$$\left. + 4\hat{\Psi}_{R.E}(2\mathbf{P}) C_f(\mathbf{P}) - 8\hat{\Psi}_{I.O}(2\mathbf{P}) D_f(\mathbf{P}) \right] \geq 0 \tag{11}$$

where we use shorthand notations

$$A_f(\mathbf{P}) = \hat{f}_{R.E}^2(\mathbf{P}) + \hat{f}_{R.O}^2(\mathbf{P}) + \hat{f}_{I.E}^2(\mathbf{P}) + \hat{f}_{I.O}^2(\mathbf{P}),$$
$$B_f(\mathbf{P}) = \hat{f}_{R.E}(\mathbf{P})\hat{f}_{R.O}(\mathbf{P}) + \hat{f}_{I.E}(\mathbf{P})\hat{f}_{I.O}(\mathbf{P}),$$
$$C_f(\mathbf{P}) = \hat{f}_{R.E}^2(\mathbf{P}) - \hat{f}_{R.O}^2(\mathbf{P}) + \hat{f}_{I.E}^2(\mathbf{P}) - \hat{f}_{I.O}^2(\mathbf{P}), \tag{12}$$
$$D_f(\mathbf{P}) = \hat{f}_{R.O}(\mathbf{P})\hat{f}_{I.E}(\mathbf{P}) - \hat{f}_{R.E}(\mathbf{P})\hat{f}_{I.O}(\mathbf{P}).$$

Inequality, Eq. (11), should be valid for any set of real functions $\hat{f}_{R.E}(\mathbf{P})$, $\hat{f}_{R.O}(\mathbf{P})$, $\hat{f}_{I.E}(\mathbf{P})$, and $\hat{f}_{I.O}(\mathbf{P})$. It is shown in Appendix 1 that the necessary and sufficient condition for this is that the integrand in Eq. (11) is non-negative for all probe functions and for all wave vectors $\mathbf{P}$.

$$\hat{\Phi}_{R.E}(\mathbf{P})A_f(\mathbf{P}) - 2\hat{\Phi}_{R.O}(\mathbf{P})B_f(\mathbf{P}) + 4\hat{\Psi}_{R.E}(2\mathbf{P})C_f(\mathbf{P}) - 8\hat{\Psi}_{I.O}(2\mathbf{P})D_f(\mathbf{P}) \geq 0 \ . \tag{13}$$

Inequality, Eq. (13) should be valid independently for every wave vector **P** and all probe functions. This allows to treat the values of the probe function Fourier components in Eq. (13) as independent real arguments. Hence, Eq. (13) can be treated as the NND condition for a four-dimensional quadratic form. Details of the standard NND analysis of corresponding matrix are presented in Appendix 2, with a final result

$$\hat{\Phi}_{R.E}(\mathbf{P}) \geq \Omega(\mathbf{P}), \quad \Omega(\mathbf{P}) \equiv \sqrt{\hat{\Phi}^2_{R.O}(\mathbf{P}) + 16|\hat{\Psi}(2\mathbf{P})|^2} \quad \text{for all } \mathbf{P}. \tag{14}$$

Several simple examples of BCFs satisfying this NND condition will be presented in Section 4 but now we present an alternative derivation of the NND condition based on the spectral analysis of the BCF.

### 3. Eigenvalues and eigenfunctions of BCF

Eigenvalues $\lambda$ and corresponding eigenfunctions $g_\lambda(\mathbf{r})$ of CF $W(\mathbf{R}, \mathbf{r})$ satisfy the integral equation

$$\iint d^2 r_1 W\left(\frac{\mathbf{r}_1 + \mathbf{r}_2}{2}, \mathbf{r}_1 - \mathbf{r}_2\right) g_\lambda(\mathbf{r}_1) = \lambda g_\lambda(\mathbf{r}_2), \tag{15}$$

and CF is NND if in only if all eigenvalues are non-negative.

We use spectral representations, of BSF, Eq. (9), and eigenfunctions

$$\hat{g}_\lambda(\mathbf{P}) = \frac{1}{4\pi^2} \iint d^2 r g_\lambda(\mathbf{r}) \exp(-i\mathbf{r} \cdot \mathbf{P}) \tag{16}$$

to simplify the integral Eq. (15) as

$$\iint d^2 P \left[ \hat{\Phi}(-\mathbf{P}) \exp(i\mathbf{P} \cdot \mathbf{r}) + 4\hat{\Psi}(-2\mathbf{P}) \exp(-i\mathbf{P} \cdot \mathbf{r}) - \frac{\lambda}{4\pi^2} \hat{g}_\lambda(\mathbf{P}) \exp(i\mathbf{P} \cdot \mathbf{r}) \right] = 0, \quad \forall \mathbf{r}. \tag{17}$$

Analytical form of Eq. (17) suggests that $\hat{g}_\lambda(\mathbf{P})$ can be sought in the form

$$\hat{g}_\lambda(\mathbf{P}) = a(\mathbf{K})\delta(\mathbf{P} - \mathbf{K}) + b(\mathbf{K})\delta(\mathbf{P} + \mathbf{K}). \tag{18}$$

Here 2-D vector **K** is the spectral parameter, which characterizes the different eigenvalues and corresponding eigenfunctions. Substitution of Eq. (18) into Eq. (17), after separating the $\exp(\pm i\mathbf{K} \cdot \mathbf{r})$ terms, lead to the following system of equations for eigenvalues $\lambda(\mathbf{K})$ and complex amplitudes $a(\mathbf{K})$ and $b(\mathbf{K})$

$$\begin{aligned} \left[\hat{\Phi}(-\mathbf{K}) - \frac{\lambda(\mathbf{K})}{4\pi^2}\right] a(\mathbf{K}) + 4\hat{\Psi}(2\mathbf{K})b(\mathbf{K}) &= 0, \\ 4\hat{\Psi}(-2\mathbf{K})a(\mathbf{K}) + \left[\hat{\Phi}(\mathbf{K}) - \frac{\lambda(\mathbf{K})}{4\pi^2}\right] b(\mathbf{K}) &= 0. \end{aligned} \tag{19}$$

In order to have non-zero eigenfunctions, this homogeneous system has to be degenerate, leading to the eigenvalues that can be presented as

$$\lambda_{1,2}(\mathbf{K}) = 4\pi^2 \left(\hat{\Phi}_{R.E}(\mathbf{K}) \pm \Omega(\mathbf{K})\right). \tag{20}$$

The NND condition, $\lambda \geq 0$, leads to the inequality, Eq. (14), confirming our earlier result. Eq. (17) suggests that there are two sets of eigenvalues corresponding to the plus and minus signs in Eq. (17). Each set is parameterized by a 2-D continuous real parameter, denoted **K** in Eq. (20). Further we will use parameter **K** to characterize the eigenfunctions, which now will be denoted as $g_{1,2}(\mathbf{r}, \mathbf{K})$

Amplitudes $a(\mathbf{K})$ and $b(\mathbf{K})$ satisfy either of equation in Eq. (19) when eigenvalues are given by Eq. (20), and eigenfunctions $g_{1,2}(\mathbf{r}, \mathbf{K})$ can be presented as

$$\begin{aligned} g_1(\mathbf{r}, \mathbf{K}) &= 4\hat{\Psi}(2\mathbf{K}) \exp(i\mathbf{r} \cdot \mathbf{K}) + \left(\hat{\Phi}_{R.O}(\mathbf{K}) + \Omega(\mathbf{K})\right) \exp(-i\mathbf{r} \cdot \mathbf{K}), \\ g_2(\mathbf{r}, \mathbf{K}) &= 4\hat{\Psi}(2\mathbf{K}) \exp(i\mathbf{r} \cdot \mathbf{K}) + \left(\hat{\Phi}_{R.O}(\mathbf{K}) - \Omega(\mathbf{K})\right) \exp(-i\mathbf{r} \cdot \mathbf{K}). \end{aligned} \tag{21}$$

It is important to notice that $\lambda_{1,2}(-\mathbf{K}) = \lambda_{1,2}(\mathbf{K})$. This does not mean however that the eigenvalues are degenerate, since $g_{1,2}(\mathbf{r},-\mathbf{K}) \propto g_{1,2}(\mathbf{r},\mathbf{K})$. This issue is related to underlying ansatz, Eq. (18), where domain for the parameter K was not specified. Turns out that it is sufficient to limit **K** to any subset **Σ** of the plane **K**, such that for any **K**, only one of two points **K** and –**K**, belongs to **Σ**. This, somewhat shoddy, argumentation is supported by the two essential properties of the eigenvalues and eigenfunctions: orthogonality and Mercer's theorem.

The inner products of eigenfunctions are calculated as

$$\int d^2 r\, g_1(\mathbf{r},\mathbf{K}_1) g_1^*(\mathbf{r},\mathbf{K}_2) = 8\pi^2 (\hat{\Phi}_{R.O}(\mathbf{K}) + \Omega(\mathbf{K}))\Omega(\mathbf{K})\delta(\mathbf{K}_1 - \mathbf{K}_2),$$
$$\int d^2 r\, g_2(\mathbf{r},\mathbf{K}_1) g_2^*(\mathbf{r},\mathbf{K}_2) = 8\pi^2 (-\hat{\Phi}_{R.O}(\mathbf{K}) + \Omega(\mathbf{K}))\Omega(\mathbf{K})\delta(\mathbf{K}_1 - \mathbf{K}_2),\quad (22)$$
$$\int d^2 r\, g_1(\mathbf{r},\mathbf{K}_1) g_2^*(\mathbf{r},\mathbf{K}_2) = 0, \quad \mathbf{K}_1, \mathbf{K}_2 \in \Sigma.$$

This allows introduction of the normalized eigenfunctions

$$\bar{g}_1(\mathbf{r},\mathbf{K}) = \frac{4\hat{\Psi}(2\mathbf{K})\exp(i\mathbf{r}\cdot\mathbf{K}) + (\Omega(\mathbf{K}) + \hat{\Phi}_{R.O}(\mathbf{K}))\exp(-i\mathbf{r}\cdot\mathbf{K})}{\sqrt{8\pi^2(\Omega(\mathbf{K}) + \hat{\Phi}_{R.O}(\mathbf{K}))\Omega(\mathbf{K})}}$$
$$\bar{g}_2(\mathbf{r},\mathbf{K}) = \frac{4\hat{\Psi}(2\mathbf{K})\exp(i\mathbf{r}\cdot\mathbf{K}) - (\Omega(\mathbf{K}) - \hat{\Phi}_{R.O}(\mathbf{K}))\exp(-i\mathbf{r}\cdot\mathbf{K})}{\sqrt{8\pi^2(\Omega(\mathbf{K}) - \hat{\Phi}_{R.O}(\mathbf{K}))\Omega(\mathbf{K})}}. \quad (23)$$

Mercer's theorem, [7], represents CF as a bilinear form in terms of eigenvalues and eigenfunctions. In our case it can be formulates as

$$\Phi(\mathbf{r}_1 - \mathbf{r}_2) + \Psi\left(\frac{\mathbf{r}_1 - \mathbf{r}_2}{2}\right) = \iint_\Sigma d^2 K \left[\lambda_1(\mathbf{K}) g_1^*(\mathbf{r}_1,\mathbf{K}) g_1(\mathbf{r}_2,\mathbf{K}) + \lambda_2(\mathbf{K}) g_2^*(\mathbf{r}_1,\mathbf{K}) g_2(\mathbf{r}_2,\mathbf{K})\right]. \quad (24)$$

It is straightforward to show that substitution of Eq. (20, 23) validates Eq. (24), and proves that ansatz, Eq. (18) captures all eigenvalues and eigenfunctions.

## 4. Examples

In this section we discuss two simple BCF examples.

### 4.1 Both BCF terms are even

In this case

$$\hat{\Phi}_{R.O}(\mathbf{P}) = 0, \quad \hat{\Psi}_{I.O}(\mathbf{p}) = 0 \quad (25)$$

and eigenvalues and eigenfunctions are

$$\lambda_{1,2}(\mathbf{K}) = 4\pi^2 (\hat{\Phi}_{R.E}(\mathbf{K}) \pm |\hat{\Psi}_{R.E}(2\mathbf{K})|) \geq 0,$$
$$\bar{g}_{1,2}(\mathbf{r},\mathbf{K}) = \frac{1}{8\pi^2}[\exp(i\mathbf{r}\cdot\mathbf{K}) \pm \exp(-i\mathbf{r}\cdot\mathbf{K})], \quad \mathbf{K} \in \Sigma. \quad (26)$$

For a simple example we cast both terms in the Gaussian form:

$$W(\mathbf{R},\mathbf{r}) = \exp\left(-\frac{r^2}{4a^2}\right) + \theta \exp\left(-\frac{R^2}{b^2}\right),$$
$$\hat{\Phi}(\mathbf{P}) = \frac{a^2}{\pi}\exp(-P^2 a^2), \quad \hat{\Psi}(\mathbf{p}) = \frac{\theta b^2}{4\pi}\exp\left(-\frac{p^2 b^2}{4}\right). \quad (27)$$

NND condition in Eq. (26) in this case is

$$a^2 \exp(-K^2 a^2) \pm \theta b^2 \exp(-K^2 b^2) \geq 0, \quad (28)$$

and Eq. (27) is a valid CF if and only if

$$b^2 \geq a^2, \quad |\theta| \leq \frac{a^2}{b^2} \leq 1. \tag{29}$$

Irradiance distribution corresponding to the BCF, Eq. (27) is a bright or dark spot on the uniform background. Parameter $\theta$ describes the contrast between the spot and background. Irradiance distribution, as well as CF, do not change on paraxial propagation. The spot width can be arbitrary small, provided that paraxial approximation is valid.

An interesting case is the critical Gaussian BCF where $b^2 = a^2$, $\theta = \pm 1$, when one of eigenvalues is identically zero, and

$$W(\mathbf{R}, \mathbf{r}) = \exp\left(-\frac{r^2}{4a^2}\right) \pm \exp\left(-\frac{R^2}{a^2}\right),$$
$$I(\mathbf{R}) = W(\mathbf{R}, 0) = 1 \pm \exp\left(-\frac{R^2}{a^2}\right) \tag{30}$$

Fig. 1 shows irradiance distributions described by Eq. (30). It is notable that irradiance of the critical Gaussian BCF is zero at coordinate origin. This suggests that each sample of corresponding random field carries an optical vortex at origin.

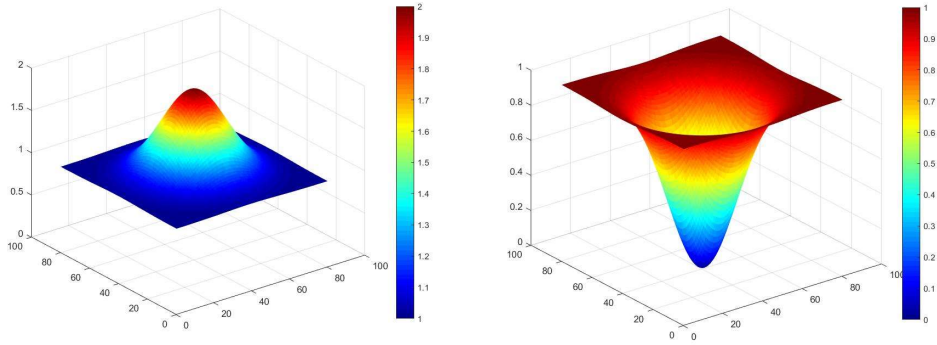

Fig. 1. Irradiance distributions for a bright, left, and dark critical Gaussian BCF beams.

### 4.2 Even/odd BCF terms

Consider case when uniform part of BCF is even, and inhomogeneous part is odd

$$W(\mathbf{R}, \mathbf{r}) = \Phi_E(\mathbf{r}) + i\Psi_O(\mathbf{R}). \tag{31}$$

In this case

$$\hat{\Phi}_{R.O}(\mathbf{P}) = 0, \quad \hat{\Psi}_{R.E}(\mathbf{p}) = 0 \tag{32}$$

and eigenvalues and eigenfunctions are

$$\lambda_{1,2}(\mathbf{K}) = 4\pi^2 \left(\hat{\Phi}_{R.E}(\mathbf{K}) \pm 4\left|\hat{\Psi}_{R.O}(2\mathbf{K})\right|\right) \geq 0,$$
$$\overline{g}_{1,2}(\mathbf{r}, \mathbf{K}) = \frac{1}{8\pi^2}\left[\exp(i\mathbf{r} \cdot \mathbf{K}) \pm \exp(-i\mathbf{r} \cdot \mathbf{K})\right], \quad \mathbf{K} \in \Sigma. \tag{33}$$

Consider a simple example

$$W(\mathbf{R}, \mathbf{r}) = \exp\left(-\frac{r^2}{4a^2}\right) + \theta\frac{X}{b}\exp\left(-\frac{R^2}{b^2}\right),$$
$$\hat{\Phi}(\mathbf{P}) = \frac{a^2}{\pi}\exp(-P^2 a^2), \quad \hat{\Psi}(\mathbf{p}) = -i\frac{\theta b^3}{8\pi}\exp\left(-\frac{p^2 b^2}{4}\right). \tag{34}$$

NND condition in Eq. (33) in this case is

$$a^2 \exp(-K^2 a^2) - \theta b^2 K_X \exp(-K^2 b^2) \geq 0, \quad \mathbf{K} = (K_X, K_Y), \tag{35}$$

It is clear from the high wave vector behavior of the left-hand part of Eq. (35) that necessarily $b^2 \geq a^2$. Further analysis, presented in Appendix 3, reveals that Eq. (35) is valid uniformly in $\mathbf{K}$ if and only if

$$|\theta| \leq \theta^* = \frac{a^2}{b^3}\sqrt{2e(b^2 - a^2)}. \tag{36}$$

The maximal irradiance contrast is achieved for $\theta = \theta^*$, Eq. (36), when irradiance is

$$I(\mathbf{R}) = 1 + \frac{a^2}{b^4}\sqrt{2e(b^2 - a^2)} X \exp\left(-\frac{R^2}{b^2}\right), \tag{37}$$

and extremal irradiance values are reached at the points $\mathbf{R}_{EX}$.

$$I(\mathbf{R}_{EX}) = 1 \pm \frac{a^2}{b^3}\sqrt{b^2 - a^2}, \quad \mathbf{R}_{EX} = \left(\pm\frac{b}{\sqrt{2}}, 0\right). \tag{38}$$

For a fixed spot size $b$ contrast can be maximized by variation of the coherence scale $a$. Namely for

$$a = \sqrt{\frac{2}{3}}b, \quad \theta^* = \sqrt{\frac{8e}{27}} \approx 0.90, \quad I(\mathbf{R}_{EX}) = 1 \pm \frac{2}{3\sqrt{3}} \approx 1 \pm 0.385. \tag{39}$$

Fig. 2 shows irradiance distribution described by Eq. (37) with parameters $a$ and $\theta$ determined by Eq. (39).

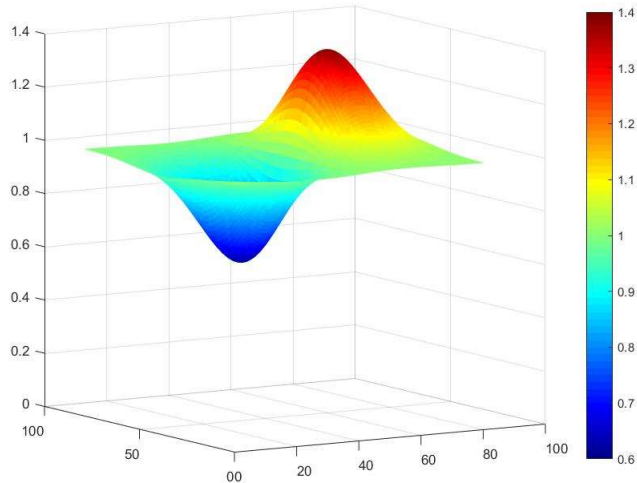

Fig. 2. Irradiance distribution for the critical even/odd Gaussian BCF, Eq. (37). Parameters $a$ and $\theta$ determined by Eq. (39) to achieve the highest possible contrast.

## 5. Discussion and conclusions

We presented a comprehensive examination of one special class of non-diffracting partially-coherent optical waves. Two-point coherence function of these waves consist of two terms, one depending only on the point separation vector, and another on the position of the central point, Eq. (1). Corresponding irradiance consist of a uniform background and a spatially confined "bump," which does not have to be positive.

We established the sufficient and necessary condition of the non-negative definiteness of BCF in terms of Fourier transforms of its terms, Eq. (14), based both on direct evaluation of the NND inequality, and on the BCF spectra.

The specific form of BCF allows for a relatively simple spectral analysis. We calculated the BCF eigenvalues that happened to have two branches of two-dimensional continuous spectra each supported at the half plane of the spectral parameter, Eq. (20). Corresponding eigen functions are combinations of the 2-D sinusoidal components.

Non-diffracting property of the BCF waves preserves the contrast between the "bump" and background in the paraxial propagation. This can lead to some practical applications in optical communication. We showed on two simple examples that the "bump" component can be positive, negative or have varying sign.

The maximum contrast is achieved for the critical BCF waves when some of the eigenvalues approach zero. We demonstrated on a simple example that irradiance of a critical BCF wave can be zero at a point, implying that every sample of such BCF field carries an optical vortex. The width of the "bump" can be arbitrary small, provided that the paraxial approximation is still valid, namely the coherence scale and the "bump" width are larger than the wavelength. However, the use of non-diffractive property of the BCF waves requires infinitely large apertures.

**Appendix 1. Proof of inequality, Eq. (13)**

**Sufficient condition.** If the integrand, Eq. (13) is non-negative for all $\mathbf{P}$, then integral, Eq. (11) is nonnegative.

**Necessary condition.** Assume that inequality, Eq. (13), is not true. Then there exists a probe function $\varphi(\mathbf{r})$ with Fourier transform $\hat{\varphi}(\mathbf{P})$ having the real, imaginary and even and odd component, as introduced in Eq. (9), such that the left-hand part of Eq. (13) is negative at least at a single point $\mathbf{P} = \mathbf{P}_0$.

$$\hat{\Phi}_{R.E}(\mathbf{P}_0)A_\varphi(\mathbf{P}_0) - 2\hat{\Phi}_{R.O}(\mathbf{P}_0)B_\varphi(\mathbf{P}_0) + 4\hat{\Psi}_{R.E}(2\mathbf{P}_0)C_\varphi(\mathbf{P}_0) - 8\hat{\Psi}_{I.O}(2\mathbf{P}_0)D_\varphi(\mathbf{P}_0) < 0. \qquad (40)$$

We introduce the auxiliary function on the wave vector space

$$\Delta_\varepsilon(\mathbf{P}) = \frac{1}{\sqrt{2\pi}\varepsilon} \exp\left(-\frac{P^2}{2\varepsilon^2}\right), \qquad (41)$$

and probe function $f_-(\mathbf{r})$ with Fourier transform

$$\hat{f}_-(\mathbf{P}) = [\hat{\varphi}_{R.E}(\mathbf{P}_0) + i\hat{\varphi}_{R.E}(\mathbf{P}_0)][\Delta_\varepsilon(\mathbf{P}-\mathbf{P}_0) + \Delta_\varepsilon(\mathbf{P}+\mathbf{P}_0)] \\
+ [\hat{\varphi}_{R.O}(\mathbf{P}_0) + i\hat{\varphi}_{R.O}(\mathbf{P}_0)][\Delta_\varepsilon(\mathbf{P}-\mathbf{P}_0) - \Delta_\varepsilon(\mathbf{P}+\mathbf{P}_0)]. \qquad (42)$$

The integrand in Eq. (11) is an even function of $\mathbf{P}$, and integration domain can be reduced to a half plane containing $\mathbf{P}_0$, but not $-\mathbf{P}_0$. The NND functional, Eq. (11) on function $\hat{f}_-(\mathbf{P})$ is calculated as

$$S[f_-(\bullet)] = (2\pi)^4 \iint d^2P \Delta_\varepsilon^2(\mathbf{P}-\mathbf{P}_0) \begin{bmatrix} \hat{\Phi}_{R.E}(\mathbf{P})A_\varphi(\mathbf{P}_0) - 2\hat{\Phi}_{R.O}(\mathbf{P})B_\varphi(\mathbf{P}_0) \\ + 4\hat{\Psi}_{R.E}(2\mathbf{P})C_\varphi(\mathbf{P}_0) - 8\hat{\Psi}_{I.O}(2\mathbf{P})D_\varphi(\mathbf{P}_0) \end{bmatrix}. \qquad (43)$$

We let $\varepsilon \to 0$, and assume that all component of the Fourier transform of the BCF are continuous functions at $\mathbf{P}_0$. Then, for example,

$$\iint d^2P \Delta_\varepsilon^2(\mathbf{P}-\mathbf{P}_0)\hat{\Phi}_{R.E}(\mathbf{P})\bigg|_{\varepsilon\to 0} = \hat{\Phi}_{R.E}(\mathbf{P}_0)\iint d^2P' \Delta_\varepsilon^2(\mathbf{P}') = \hat{\Phi}_{R.E}(\mathbf{P}_0), \qquad (44)$$

and

$$S[f_-(\bullet)] = (2\pi)^4 \begin{bmatrix} \hat{\Phi}_{R.E}(\mathbf{P}_0)A_\varphi(\mathbf{P}_0) - 2\hat{\Phi}_{R.O}(\mathbf{P}_0)B_\varphi(\mathbf{P}_0) \\ + 4\hat{\Psi}_{R.E}(2\mathbf{P}_0)C_\varphi(\mathbf{P}_0) - 8\hat{\Psi}_{I.O}(2\mathbf{P}_0)D_\varphi(\mathbf{P}_0) \end{bmatrix} < 0, \qquad (45)$$

as follows from Eq. (43). This proves that Eq. (13) is a necessary condition for the NND of the BCF.

**Appendix 2. Analysis of the quadratic form, Eq. (13).**

In order to shorten the following equations we introduce the shorthand notations for this Appendix only.

$$\hat{f}_{R.E}(\mathbf{P}) = x, \ \hat{f}_{R.O}(\mathbf{P}) = y, \ \hat{f}_{I.E}(\mathbf{P}) = u, \ \hat{f}_{I.O}(\mathbf{P}) = v, \\
\hat{\Phi}_{R.E}(\mathbf{P}) = l, \ \hat{\Phi}_{R.O}(\mathbf{P}) = m, \ 4\hat{\Psi}_{I.E}(2\mathbf{P}) = n, \ 4\hat{\Psi}_{I.O}(2\mathbf{P}) = h. \qquad (46)$$

After that, the NND inequality, Eq. (13) is presented as

$$l(x^2 + y^2 + u^2 + v^2) - 2m(xy + uv) + n(x^2 - y^2 + u^2 - v^2) - 2h(yu - xv) \geq 0. \qquad (47)$$

Real symmetric matrix of this quadratic form is

$$\begin{pmatrix} x \\ y \\ u \\ v \end{pmatrix}^T \begin{pmatrix} l+n & -m & 0 & h \\ -m & l-n & -h & 0 \\ 0 & -h & l+n & -m \\ h & 0 & -m & l-n \end{pmatrix} \begin{pmatrix} x \\ y \\ u \\ v \end{pmatrix}. \qquad (48)$$

Characteristic equation for this matrix is readily calculated as

$$\left[(l-\mu)^2 - (m^2+n^2+h^2)\right] = 0, \qquad (49)$$

and the NND condition for quadratic form, Eq. (47) is

$$\mu = l \pm \sqrt{m^2+n^2+h^2} \geq 0, \qquad (50)$$

After returning the original notations, Eq. (14) is recovered.

## Appendix 3. Derivation of the NND condition for the Even-Odd BCF example.

We rewrite inequality, Eq. (35) as

$$\exp\left[(b^2-a^2)(K_X^2+K_Y^2)\right] \geq \theta \frac{b^3}{a^2} K_X. \qquad (51)$$

Since $b^2 > a^2$, the most restrictive case at the wave vector plane is $K_Y = 0$. We are interested in the maximum possible value of parameter $\theta = \theta^*(a,b)$, such that the inequality

$$\exp\left[(b^2-a^2)K_X^2\right] - \theta \frac{b^3}{a^2} K_X \geq 0 \qquad (52)$$

is satisfied for any $K_X$, and for given $a$ and $b$. This implies that at the point $K_X = K_X^*(a,b)$, where the left-hand part of Eq. (52) reaches its minimum in $K_X$, inequality in Eq. (52) turns into equality. This leads to the system of equations for $\theta^*$ and $K_X^*$

$$\exp\left[(b^2-a^2)(K_X^*)^2\right] = \theta^* \frac{b^3}{a^2} K_X^*,$$
$$\frac{\partial}{\partial K_X} \exp\left[(b^2-a^2)(K_X^*)^2\right] = \frac{\partial}{\partial K_X} \theta^* \frac{b^3}{a^2} K_X^*. \qquad (53)$$

The solution is readily calculated as

$$K_X^* = \sqrt{\frac{1}{2(b^2-a^2)}}, \quad \theta^* = \frac{a^2}{b^3}\sqrt{2e(b^2-a^2)}, \qquad (54)$$

Where $e$ is the Euler's number. The last equation leads to the NND condition, Eq. (36)

**References**
1. L. Mandel and E. Wolf, *Optical Coherence and Quantum Optics* (Cambridge U. Press, 1995).
2. S. A. Ponomarenko, W. Huang and M. Cada, "Dark and antidark diffraction-free beams," Opt. Lett. **32**(17), 2508–2510 (2007).
3. X. Li, S. A. Ponomarenko, Z. Xu, F. Wang, Y. Cai and C. Liang, "Universal self-similar asymptotic behavior of optical bump spreading in random medium atop incoherent background," Opt. Lett. **45**(3), 698–701 (2020).
4. M. Charnotskii, "Universal self-similar asymptotic behavior of optical bump spreading in random medium atop incoherent background: comment," Opt. Lett. **45**(13), 3510 (2020).
5. X. Li, S. A. Ponomarenko, Z. Xu, F. Wang, Y. Cai and C. Liang, "Universal self-similar asymptotic behavior of optical bump spreading in random medium atop incoherent background: reply," Opt. Lett. **45**(13), 3511 (2020).
6. F. Riesz and B. Sz.-Nagy, Functional Analysis (Dover, 1990).
7. R. Courant and D. Hilbert, *Methods of Mathematical Physics*, Vol 1, (Interscience, 1953).